\documentclass[prd,amsmath,amssymb]{revtex4}
\usepackage{graphicx}
\usepackage{dcolumn}
\usepackage{bm}

\begin{document}

\vspace*{1.5cm}

\title{A Study of the
Leaky Pipeline Phenomenon\\ for Women in Physics
Past the Postdoctoral Level, \\ and a Critique of the AIP 
2005 report on Women in Physics and Astronomy}

\author{S.Towers, PhD\\ Physics Department\\ State University of New York at Stony Brook \vspace*{0.5cm}}
 
\date{\today}
\begin{abstract}
\vspace*{3.0cm}

The author has recently examined the departmental web pages of 
the `top 50' physics research universities, as ranked by the National Research 
Council (NRC) \cite{bib:nrc}.  
Most of the departmental web pages
contained biographical data (ie; year and institute of PhD, etc) of their
faculty members.  Of the approximately 1750 faculty 
members at the `top 50' universities that
were examined, approximately 100 were female, and around 
1425 had available biographical data.

Based on this data, the {\bf predicted} fractions of female faculty
members at the `top 50' universities are
$0.129$,$0.104$, and $0.052$ at the assistant, associate, and
full faculty levels, respectively.  

The {\bf observed} fractions are
$0.107$, $0.109$, $0.043$, respectively.  
 
The overall observed number of
women faculty
is about 15\% less than expected, and the depletion is
statistically significant. 

Unfortunately, the study finds that 
the "leaky pipeline" is found to be alive and well for women
in academic physics above the postdoctoral level, at all stages of the
faculty career ladder.
This result is stark contrast with the conclusion of the American
Institute of Physics (AIP)
2005 report on Women in Physics and Astronomy; the AIP report concludes that
women are actually {\bf more} likely to be hired at the faculty level than their
male peers.

In this paper, we will 
discuss the two key flaws in the AIP analysis that led to
their faulty conclusion, then describe in detail the
analysis performed by the author that
corrects these flaws to get an accurate estimate of the
`leakiness' of the academic pipeline for women 
physicists past the
postdoctoral level.

\vspace*{2.0cm}
\centerline{\em To be submitted to the American Journal of Physics}
\end{abstract}


\maketitle
\newpage
 
Many studies have documented the "leaky pipeline" phenomenon
for women in the academic hard sciences (see, for instance, references \cite{bib:berry} and 
\cite{bib:nsf}).
A recent report that caught the author's interest was the American
Institute of Physics 2005 report on Women in Physics and Astronomy \cite{bib:aipreport};
this report is rather unique in that it concludes that the leaky pipeline
phenomenon does not exist past the doctoral level. In fact,
the report concludes 
that the observed fraction of female faculty members
is actually {\bf higher} than expected.

However, careful examination of the report reveals that the analysis
that led to this conclusion was
flawed; first, the report lumps faculty members at PhD granting
universities together with faculty members at teaching colleges.
At teaching colleges,
the faculty are more likely to be female, yet much
less likely to teach physics to physicists.
Thus the AIP report gives little indication of the fraction
of women at the faculty level at the universities
in America that produce the majority of physics doctoral degrees.
Second, the analysis performed for the report did not
properly take into account the differing age distributions of
male and female faculty members (the report used instead
the combined age distribution of both males and females, which of
course is completely dominated by the males, since they
constitute over 90\% of the sample).  
This flaw has a significant
effect on the predicted fraction of female faculty
(the predicted fraction goes up when the analysis is performed
taking the differing age distributions properly into account).
 
To perform a detailed analysis that corrects both of these
problems, the author began with an examination of the
departmental web pages of the `top 50'
physics universities, as ranked by the National Research
Council (NRC) \cite{bib:nrc}.  These universities produce
the majority of BSc's and PhD's in America.
Most of the departmental web pages
contained biographical data (ie; year and institute of PhD, etc) of their
faculty members.  
In this study astronomers were excluded because they have a different
fraction of women participating in the field than other areas
of physics.  Adjunct, visiting, research, and affiliated professors were also excluded.
The study also excluded faculty members who had received
their degree from a non-American institution.
 
Of 1743 faculty members at the `top 50' universities that
were ultimately examined, 101 were female, 
and a total of 1425 had available biographical data.

\begin{figure}[h]
\includegraphics[scale=0.5]{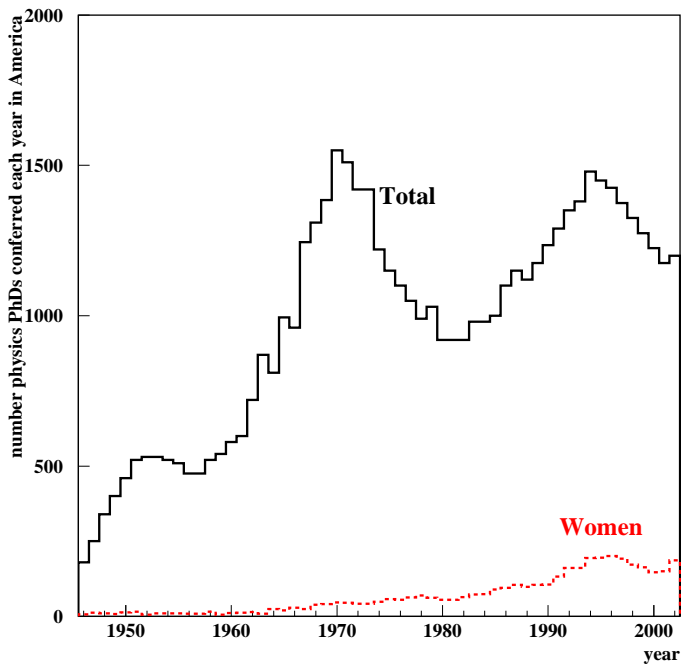}
\caption{
\label{fig:phd}
Number of physics PhD graduates produced in America each year.
}
\end{figure}

To obtain the predicted fraction of female professors
from this data, we begin with the number
of PhD's granted each year to both males and females in America 
\cite{bib:aip} (see Figure \ref{fig:phd}).
We then work out the probability that a male in a particular PhD graduating class
will be a professor at one of the `top 50' universities in 2005;  
we do this
by dividing 
the year-of-PhD distribution of male professors with the distribution
of the number of male PhD's graduating each year.
If the leaky pipeline does not exist, the female `be-a-professor-in-2005'
probability will be exactly the same
as for the males in each graduating class. 

In this manner, we obtain a
prediction of the number of female professors we expect to see at the `top 50' 
universities in 2005.

\begin{figure}
\includegraphics[scale=0.5]{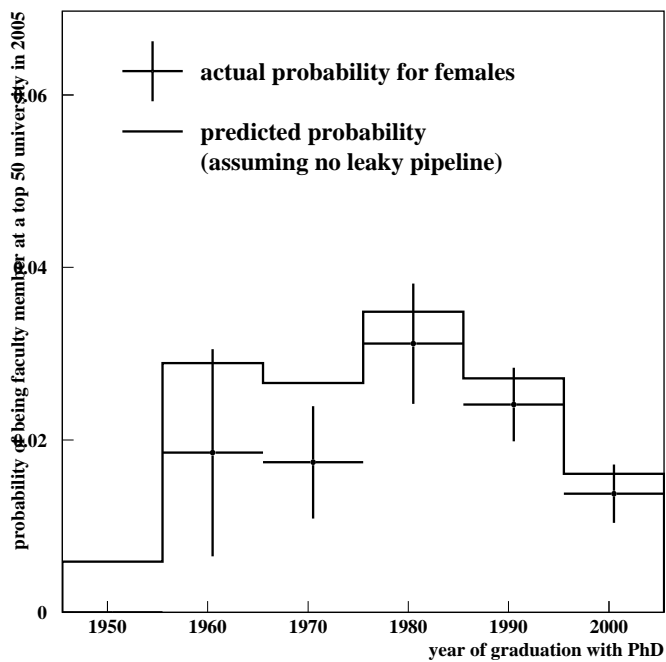}
\caption{
\label{fig:prob}
Probability that a female who graduated in a particular year will be a 
physics professor
in 2005 at one of the `top 50' American physics
universities.  Points are the actual distribution, and the histogram indicates
the predicted distribution, obtained assuming that females and males from
the same graduating class have
the same relative probability of being a professor in 2005.
}
\end{figure}

Figure~\ref{fig:prob} shows the actual probability versus year-of-PhD that
a female physicist will be a faculty member at one of the `top 50'
universities in 2005.  The histogram indicates the predicted distribution,
obtained assuming that females 
have
the same relative probability of being a professor in 2005 
as males from
the same graduating class.
The actual distribution is systematically 
lower than the predicted distribution.

\begin{figure}
\includegraphics[scale=0.5]{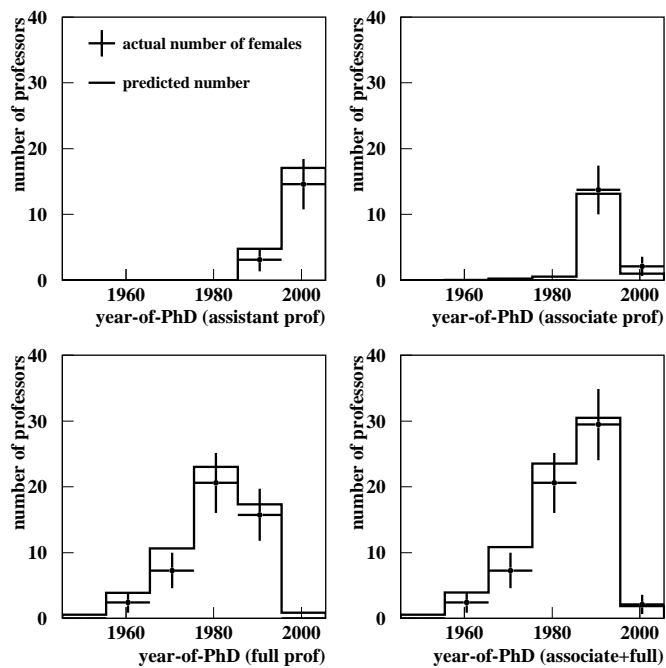}
\caption{
\label{fig:all}
Year-of-PhD of female professors in 2005 (points), for assistant, associate
and full professors.  The histograms indicate the predicted year-of-PhD
distributions,
obtained assuming that females and males from
the same graduating class have
the same relative probability of being a professor in 2005.
}
\end{figure}

\begin{figure}
\includegraphics[scale=0.5]{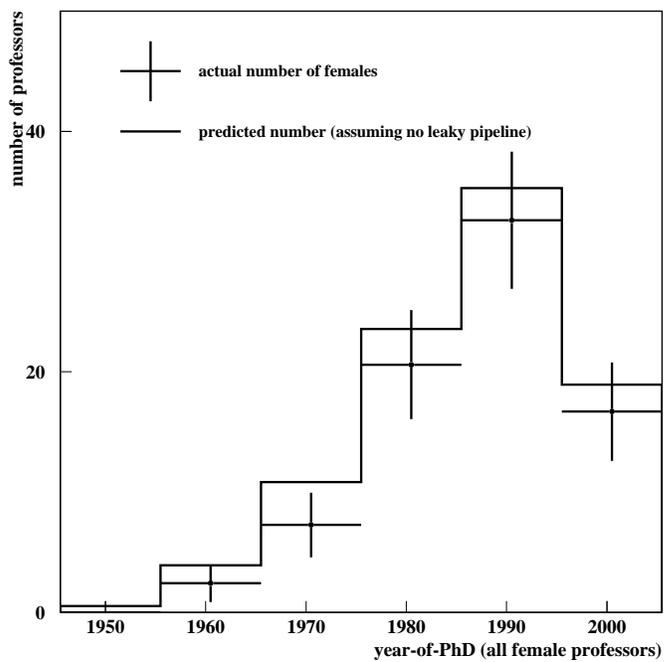}
\caption{
\label{fig:summary}
Year-of-PhD of all female professors in 2005 (points).
The histogram indicates the predicted year-of-PhD
distribution,
obtained assuming that females and males from
the same graduating class have
the same relative probability of being a professor in 2005.
}
\end{figure}

Figure~\ref{fig:all} shows the year-of-PhD distributions of
female assistant, associate, and full professors.  The histograms
again indicate
the predicted distribution,
obtained assuming that females and males from
the same graduating class have
the same relative probability of being a professor in 2005.
Figure~\ref{fig:summary} shows the year-of-PhD distributions of
all female professors.
Every point in the actual distribution is lower than the predicted.

Based on this data, the {\bf predicted} fractions of female faculty
members at the `top 50' universities  are
$0.129$,$0.104$, and $0.052$ at the assistant, associate, and
full faculty levels, respectively.  The {\bf observed} fractions are
$0.107$, $0.109$, $0.043$, respectively.  
 
It is interesting to note that the fraction of female associate professors
is actually higher than predicted.  However, Figure~\ref{fig:summary}
shows an overall depletion of professors for years-of-PhD
1984 and onwards.  It thus appears that the excess may well be due to women
languishing longer at the associate professorship level than their
male peers (ie; the excess probably reflects a `clog' in the
academic pipeline at the associate professor level).

The overall observed number of
women faculty
is about 15\% less than expected, and the depletion is
statistically significant at a level of $1.7$ standard deviations. 

\pagebreak
\section{Conclusion}
This study finds the leaky pipeline phenomenon exists for women past
the postdoctoral level at a level of around 15\%.  Some may wonder if this
is a big enough leak to be a problem.  In human terms, however, a
15\% leak means that we are missing one out of six
women who, in an equitable society,
would have been physics faculty members.  There are so few women at the faculty
level in physics, that losing one out of every six is in fact a serious
concern.  Especially if we think about what it must take to convince
someone to leave a field when, by that point in their careers, they
have committed their working lives to physics, and have gone
through at least a decade of higher education to get there.

Not all women who become physics faculty members have experienced gender
discrimination during their careers.  However, many do, and it is
unfortunate that the combination of gender discrimination
and a `glass ceiling' phenomenon in the field is preventing more women
from becoming physics faculty members.

The author is an experimental particle physicist,
and has observed over the years the serious obstacles that
her female colleagues have had to face as they try to advance in the
field.  There is indeed widespread discrimination against many women
in physics, and women with children seem to be particularly vulnerable;
for instance, the author is personally acquainted with three female
physicists who, after
having children, had to work for free or
a substantially reduced rate compared to their peers, simply
to remain in the field.  The only other choice available to
them was to simply drop out of the academic pipeline all together.
Conversely, the author knows literally hundreds of
male physicists past the doctoral level, but is not aware of a single
male who has had their pay cut off or substantially reduced for any reason.

Given some of the chilling incidents of discrimination against
females that the author has personally
observed to transpire within physics academia, it is somewhat surprising that
the relative leak of females
in the academic pipeline past the postdoctoral level
is {\bf only} 15\%.  
A male colleague once mentioned to the author
that he felt very sorry for many of his
female colleagues in physics because he felt the message that they were
persistently
given was that `Yes! the good news is that
you {\bf can} succeed in physics as a female! (you
just need to be prepared to chew your own leg off to do so)'.
 
Tragically, some of the women who have ultimately
made it to the faculty level likely did have to `chew their own leg
off' to get there.
Even more tragically, their stories are almost never told because they
(quite rightly) fear repercussions to their career if they speak out.
The chilly climate that removes one out of every six female potential physics
faculty members
needs to be changed if particpation of women
in physics is to be increased at all levels.

\pagebreak

\end{document}